\documentclass[5p]{elsarticle}
\pdfoutput=1
\usepackage{amsmath}
\usepackage{amsfonts}
\usepackage{array,multirow}
\usepackage{booktabs}
\usepackage{bm}
\usepackage{graphicx}
\usepackage{subcaption}
\usepackage{float}

\usepackage{xcolor}
\journal{Journal of \LaTeX\ Templates}

\begin{document}

\begin{frontmatter}

\title{Recurrent Inference Machines as inverse problem solvers for MR relaxometry}

\author[1]{E. R. Sabidussi}
\author[1]{S. Klein}
\author[2]{M. W. A. Caan}
\author[3]{S. Bazrafkan}
\author[3]{A. J. den Dekker}
\author[3]{J. Sijbers}
\author[1,4]{W. J. Niessen}
\author[1]{D. H. J. Poot}

\address[1]{Erasmus MC University Medical Center, Department of Radiology and Nuclear Medicine, Rotterdam, The Netherlands}
\address[2]{Amsterdam UMC, Biomedical Engineering and Physics, University of Amsterdam, Amsterdam, The Netherlands}
\address[3]{imec-Vision Lab, Department of Physics, University of Antwerp, Antwerp, Belgium}
\address[4]{Delft University of Technology, Delft, the Netherlands}



\begin{abstract}
In this paper, we propose the use of Recurrent Inference Machines (RIMs) to perform $T_1$ and $T_2$ mapping. The RIM is a neural network framework that learns an iterative inference process based on the signal model, similar to conventional statistical methods for quantitative MRI (QMRI), such as the Maximum Likelihood Estimator (MLE). This framework combines the advantages of both data-driven and model-based methods, and, we hypothesize, is a promising tool for QMRI. Previously, RIMs were used to solve linear inverse reconstruction problems. Here, we show that they can also be used to optimize non-linear problems and estimate relaxometry maps with high precision and accuracy. The developed RIM framework is evaluated in terms of accuracy and precision and compared to an MLE method and an implementation of the ResNet. The results show that the RIM improves the quality of estimates compared to the other techniques in Monte Carlo experiments with simulated data, test-retest analysis of a system phantom, and in-vivo scans. Additionally, inference with the RIM is 150 times faster than the MLE, and robustness to (slight) variations of scanning parameters is demonstrated. Hence, the RIM is a promising and flexible method for QMRI. Coupled with an open-source training data generation tool, it presents a compelling alternative to previous methods.

\end{abstract}

\begin{keyword}
Quantitative MRI\sep Relaxometry \sep Deep learning \sep Mapping \sep Recurrent inference machines
\end{keyword}

\end{frontmatter}

\section{Introduction}

MR relaxometry is a technique used to measure intrinsic tissue properties, such as $T_1$ and $T_2$ relaxation times. Compared to qualitative weighted images, quantitative $T_1$ and $T_2$ maps are much less dependent on variations of hardware, acquisition settings, and operator \citep{cercignani_2018}. Additionally, because measured  $T_1$ and $T_2$ maps are more tissue-specific than weighted images, they are promising biomarkers for a range of diseases \citep{cheng_2012, conlon_1988, erkinjuntti_1987, larsson_1989, lu_2019}. \par

Thanks to their low dependence on hardware and scanning parameters, quantitative maps are highly reproducible across scanners and patients \citep{Weiskopf2013}, presenting variability comparable to test-retest experiments within a single center \citep{deoni_2008}. The low variability allows for direct comparison of tissue properties between patients and across time \citep{cercignani_2018}. However, to ensure that quantitative maps are reproducible, mapping methods must produce estimates with low variance and bias. \par

Conventionally, quantitative maps are estimated by fitting a known signal model to every voxel of a series of weighted images with varying contrast settings. The Maximum Likelihood Estimator (MLE) is a popular statistical method used to estimate parameters of a probability density by maximizing the likelihood that a signal model explains the observed data and is extensively used in quantitative mapping \citep{ramos-llorden_2017, smit_2013, sijbers_2004}. Usually, MLE methods estimate parameters independently for each voxel. This may lead to high variability for low SNR scans. Spatial regularization can be added to the MLE (referred to as the Maximum a Posteriori - MAP) to enforce spatial smoothness, but demands high domain expertize. Additionally, for most signal models, MLE/MAP methods require an iterative non-linear optimization, which is relatively slow for clinical applications and might demand complex algorithm development. \par

Despite the current success of deep learning methods in the medical field, their application to Quantitative MRI (QMRI) is still affected by the lack of large in-vivo training sets. Specifically in MR relaxometry, the use of neural networks is still limited. Previous works successfully applied deep learning in cardiac MRI \citep{Jeelani_2020} and knee \citep{liu_2019}, but they required the scans of many subjects to train the networks and were dependent on alternative mapping methods to generate training labels. This limitation was addressed in \citet{cai_2018} and \citet{Shao_2020} by using the Bloch equations to generate simulated data to train convolutional neural networks in $T_1$ and $T_2$ mapping. However, estimation precision, a central metric in QMRI, was not reported. It is unclear, therefore, how well these methods would perform with noisy in-vivo data.\par

In this paper, we propose a new framework for MR relaxometry based on the Recurrent Inference Machines (RIMs) \citep{putzky_2017}. RIMs employs a recurrent convolutional neural network (CNN) architecture and, unlike most CNNs, learns a parameter inference method that uses the signal model, rather than a direct mapping between input signal and estimates. This hybrid framework combines the advantages of both data-driven and model-based methods, and, we hypothesize, is a promising tool for QMRI. \par

Previously, RIMs were used to solve linear inverse problems to reconstruct undersampled MR images \citep{Lonning_2019} and radio astronomy images \citep{Morningstar_2019}. In both works, synthetic, corrupted training signals (i.e. images) were generated from high-quality image labels using the forward model. \par

A significant limitation on the use of deep learning in MR relaxometry is the lack of large publicly available datasets. The acquisition of in-vivo data is a costly and time consuming process, limiting the size of training datasets and reducing flexibility in terms of the pulse sequence and scanning parameters. Using model-based strategy for data generation (in contrast to costly acquisitions) allows the creation of arbitrarily large training sets, where observational effects (e.g., acquisition noise, undersampling masks) and fixed model parameters are drawn from random distributions. This represents an essential advantage over other methods that rely entirely on acquired data. Yet, the lack of high-quality training labels (i.e. ground-truth $T_1$ and $T_2$ maps) limits the variability of training signals. Here, we also generate synthetic training labels to achieve sufficient variation in the training set.\par

We compared the proposed framework with an MLE method and an implementation of the ResNet as a baseline for conventional deep learning QMRI methods. In contrast to MLE methods with user-defined prior distribution to enforce tissue smoothness, the RIM learns the relationship between neighboring voxels directly from the data, making no assumptions about the prior distribution of values. This might improve mapping robustness to acquisition noise. \par

We evaluated each method in terms of the precision and accuracy of measurements. First, noise robustness was assessed via Monte Carlo experiments with a simulated dataset with varying noise levels. Second, we evaluated the quantitative maps' quality concerning each method's ability to retain small structures within the brain. Third, the precision and accuracy in real scans were evaluated via a test-retest experiment using a hardware phantom. Lastly, we used in-vivo scans to evaluate precision in a test-retest experiment with two healthy volunteers.

\section{QMRI framework}
\subsection{Signal modeling}

Let $\bm{\kappa}$ be the parameter maps to be inferred, such that $\bm{\kappa}(\bm{x}) \in \mathbb{R}^{Q}$ is a vector containing $Q$ tissue parameters of a voxel indexed by the spatial coordinate $\bm{x} \in \mathbb{N}^{D}$. Then, we assume that the MRI signal in each voxel of a series of $N$ weighted images $\bm{S} = \{S_1, ..., S_N\}$ follows a parametric model $f_n(\bm{\kappa}(\bm{x})): \mathbb{R}^{Q} \mapsto \mathbb{R}$ so 
\begin{align} 
	S_n(\bm{x}) = f_n(\bm{\kappa}(\bm{x})) + \epsilon(\bm{x}), \label{eq1}
\end{align}
where $\epsilon(\bm{x})$ is the noise at position $\bm{x}$. \par

For images with signal-to-noise ratio (SNR) larger than three, the acquired signal at position $\bm{x}$ can be well described by a Gaussian distribution \citep{Jan_Sijbers_2, gudbjartsson_1995}, with probability density function denoted by $p(S_n(\bm{x}_m)\ |\ f_n(\bm{\kappa}(\bm{x}_m)), \ \sigma)$, where $m \in \{1,...,M\}$ is the voxel index, $M$ the number of voxels within the MR field-of-view and $\sigma$ is the standard deviation of the noise. \par

\subsection{Quantitative mapping}
\subsubsection{Regularized Maximum Likelihood Estimator}

The Maximum Likelihood Estimator (MLE) is a statistical method that infers parameters of a model by maximizing the likelihood that the model explains the observed data. Because the MLE is asymptotically unbiased and efficient (it reaches the Cramér-Rao lower bound for a large number of weighted images) \citep{swamy_1971}, it was chosen as the reference method for this study. \par

Assume $P(\bm{S}|\bm{f}(\bm{\kappa}) , \sigma)$ is the joint PDF of all independent voxels in $\bm{S}$ from which a negative log-likelihood function $L(\bm{\kappa}, \sigma | \bm{S})$ is defined. Additionally, let $\Psi(\bm{\kappa})$ be the $\log$ of a prior probability distribution over $\bm{\kappa}$, introduced to enforce map smoothness. Then the ML estimates $\bm{\hat{\kappa}}$ are found by solving
\begin{align} 
	\hat{\bm{\kappa}} = \arg \min _{\bm{\kappa}} L(\bm{\kappa}, \sigma | \bm{S}) + \Psi(\bm{\kappa}), \label{eq2}
\end{align}
in which we assume that $\sigma$ can be estimated by alternative methods and is, therefore, not optimized. \par

Note that, although Eq.\ref {eq2} strictly defines an MAP estimator, we choose to use the term regularized MLE to emphasize that $\Psi(\bm{\kappa})$ is only applied to promote maps that vary slowly in space. In this work, regularization is used to encourage spatial smoothness of the inversion efficiency map (i.e. $B_1$ inhomogeneity), while maps linked to proton density and tissue relaxation times are not regularized and their estimation occurs exclusively at the voxel level. Herein, we refer to this method simply as MLE.\par

\subsubsection{ResNet}

The Residual Neural Network (ResNet) is a type of feed-forward network that learns to directly map input data to training labels using a concatenation of convolutional layers. It was developed by \cite{he_2016} as a solution to the degradation problem that emerges when building deep models \citep{he_2014}. Skip connections between layers of the network allow the ResNet to fit to the residual of the signal, rather than to the original input, making identity learning simpler, and ensuring that a deeper network will not perform worse than its shallower counterpart in terms of training accuracy \citep{he_2016}. For that reason, and because it was shown to be a suitable method for QMRI \citep{cai_2018}, we chose the ResNet as the reference deep learning method for this study. \par

Let $\Lambda_{\phi}: \mathbb{R}^N \mapsto \mathbb{R}^Q$ represent a ResNet model for QMRI, parameterized by $\phi$, that maps the acquired signal $\bm{S}$ to tissue parameters $\bm{\kappa}$, specifically $\bm{\hat{\kappa}} = \Lambda_{\phi}(\bm{S})$. The learning task is to find a model $\Lambda_{\hat{\phi}}$ such that the difference between $\bm{\hat{\kappa}}$ and $\bm{\kappa}$ is minimal in the training set, that is
\begin{align} 
	\hat{\phi} = \arg \min_{\phi} \left\|\bm{\kappa} - \Lambda_{\phi}(\bm{S})\right\|_{2}^{2}  \label{eq4}.
\end{align}

\section{The Recurrent Inference Machine: a new framework for QMRI}
In the context of inference learning \citep{Chen_2015, Zheng_2015}, the Recurrent Inference Machine (RIM) \citep{putzky_2017} framework was conceived to mitigate limitations linked to the choice of priors and optimization strategy. By making them implicit within the network parameters, the RIM jointly learns a prior distribution of parameters and the inference model, unburdening us from selecting them among a myriad of choices.  \par
With this framework, Eq.\ref{eq2} is solved iteratively, in an analogous way to a regularized gradient-based optimization method. The RIM uses the gradients of the likelihood function to enforce the consistency of the data and to plan efficient parameter updates, speeding up the inference process. Additionally, because this framework is based on a convolutional neural network, it learns and exploits the neighborhood context, providing an advantage over voxel-wise methods. Note that, rather than explicitly evaluating $\Psi(\bm{\kappa})$, the RIM learns it implicitly from the labels in the training dataset. 

\begin{figure}[ht!]
\vspace{9pt}
\centering
\begin{subfigure}{0.4\textwidth}
\includegraphics[trim={0.1cm 0.1cm 0.1cm 0.1cm}, clip, width=0.99\linewidth]{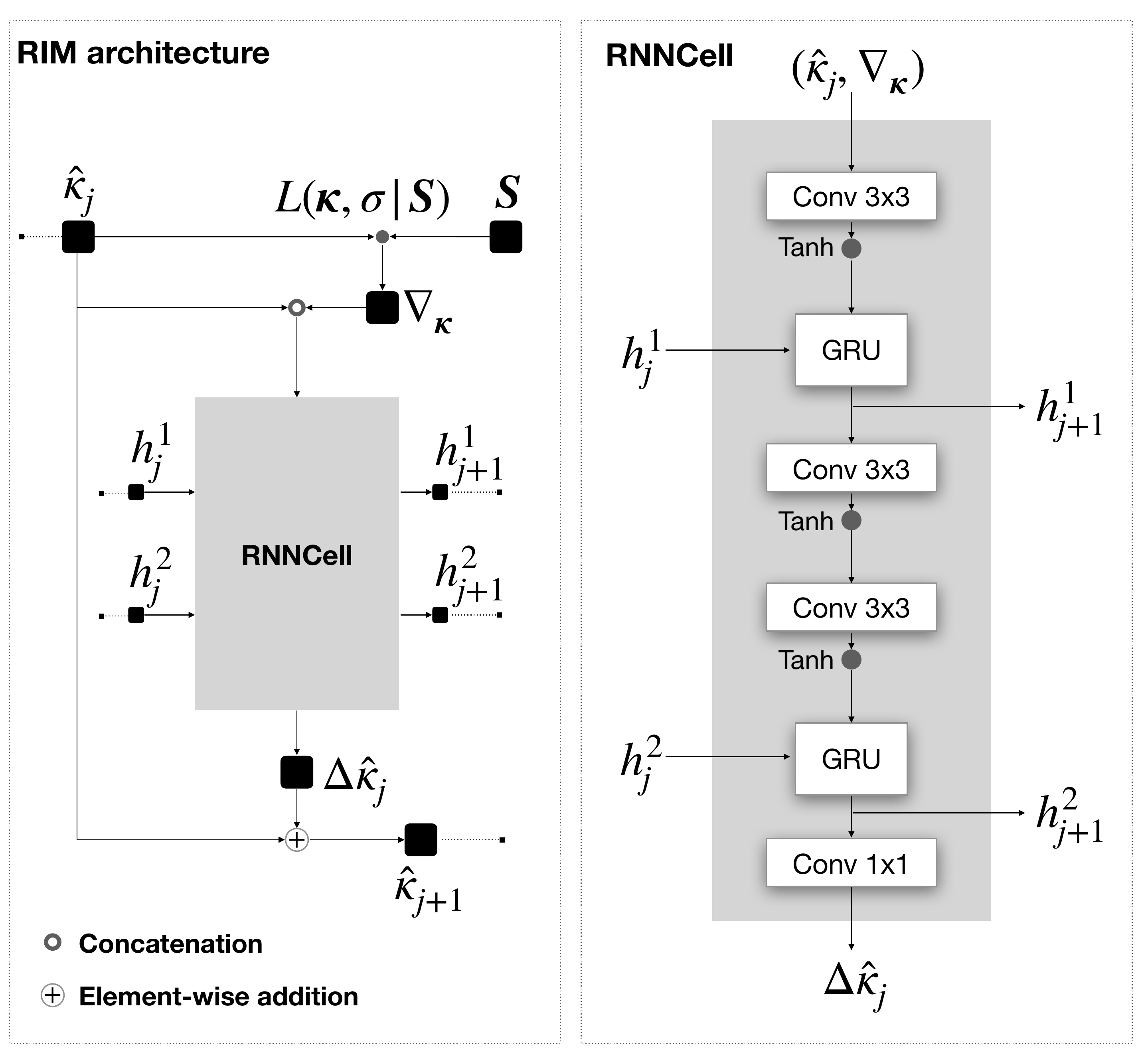}
\caption{Recurrent Inference Machine framework}
\label{fig:rim_architecture}
\end{subfigure}%

\begin{subfigure}{0.4\textwidth}
\vspace{9pt}
\centering
\includegraphics[trim={0.1cm 0.1cm 0.1cm 0.1cm}, clip, width=0.99\linewidth]{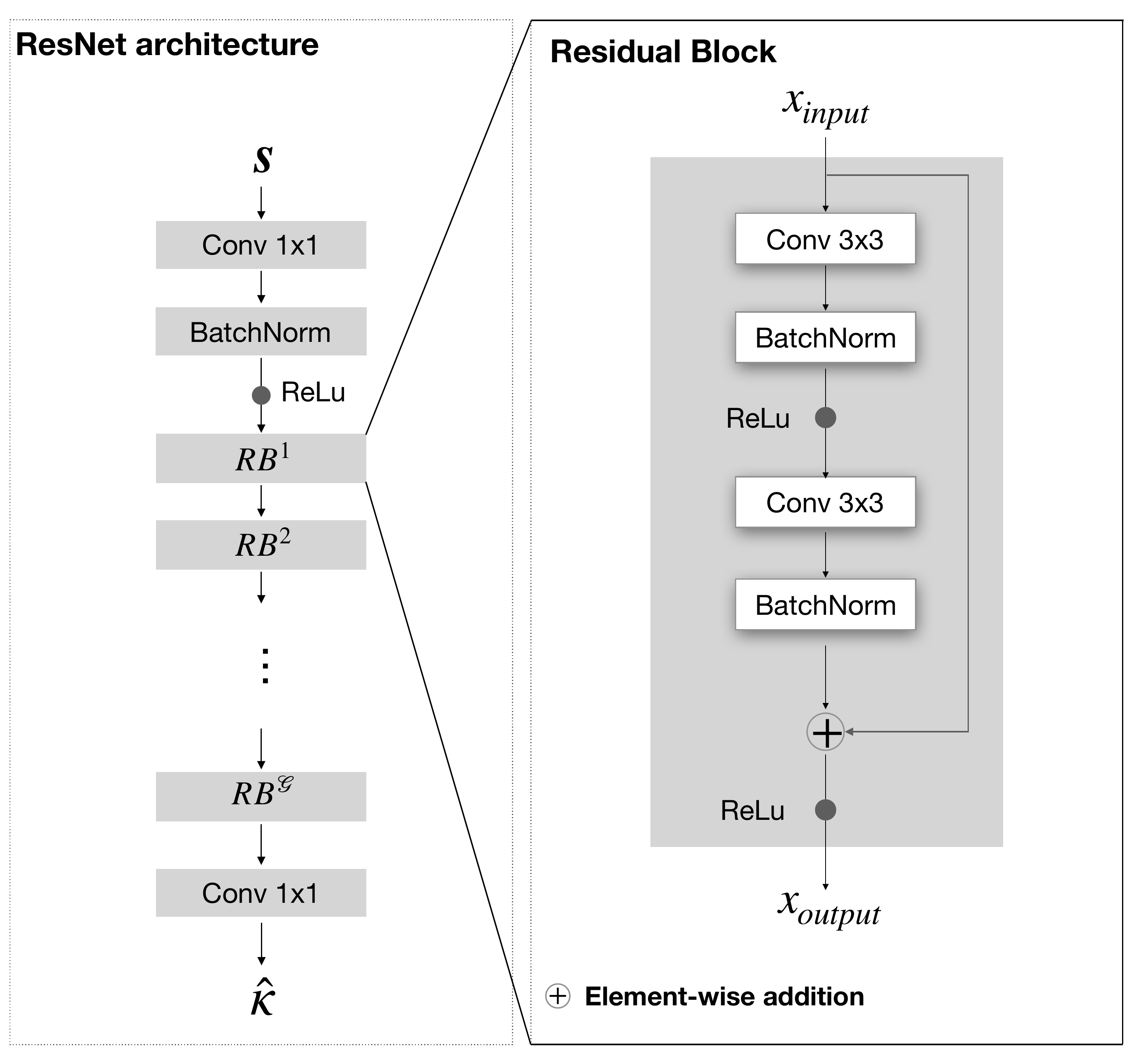}
\caption{Residual Neural Network architecture}
\label{fig:resnet_architecture}
\end{subfigure}%
\caption{a) The RIM architecture in detail. The general RIM framework is shown on the left. Dashed lines indicate information passed through iterations. The RNNCell detail is shown on the right of a). Memory states $h^*_{j+1}$ are passed to the next iteration step and used within the Gated Recurrent Units (GRU) to control the relevant information to be used from previous iterations. b) The ResNet architecture is composed of a concatenation of $\mathcal{G}$ residual blocks.}
\end{figure}

At a given optimization step $j \in \{0,...,J-1\}$, the RIM receives as input the current estimate of parameters, $\bm{\hat{\kappa}}_j$, the gradient of the negative log-likelihood $L$ with respect to $\bm{\kappa}$, $\nabla_{\bm{\kappa}}$, and a vector of memory states $\bm{h}_j$ the RIM can use to keep track of optimization progress and perform more efficient updates. The network outputs an update to the current estimate and the memory state to be used in the next iteration. The update equations for this method are given by
\begin{align} 
	\{ \Delta\bm{\hat{\kappa}}_{j+1}, \bm{h}_{j+1} \} &= \bm{g}_\gamma(\bm{\hat{\kappa}}_{j}, \ \nabla_{\bm{\kappa}}, \ \bm{h}_j), \label{eq5a}
	\\
	\bm{\hat{\kappa}}_{j+1} &= \bm{\hat{\kappa}}_{j} + \Delta\bm{\hat{\kappa}}_{j+1}, \label{eq5b}
\end{align}
where $\Delta\bm{\hat{\kappa}}_{j+1}$ is the output of the network and denotes the incremental update to the estimated maps at optimization step $j+1$ and $\bm{g}_\gamma$ represents the neural network portion of the framework, called RNNCell, parameterized by $\gamma$. A diagram of the RIM is shown on the left of Fig. \ref{fig:rim_architecture}. \par

Predictions are compared to a known ground-truth and losses are accumulated at each step, with total loss given by
\begin{align} 
	\hat{\gamma} = \arg \min_{\gamma} \frac{1}{J} \sum_{j=0}^{J-1} \left\|\bm{\kappa} - \bm{\hat{\kappa}}_{j+1}  \right\|_{2}^{2}  \label{eq6}
\end{align}
where $J$ is the total number of optimization steps and $\hat{\gamma}$ is the optimal inference model given the training data. \par

It is important to notice that the RIM uses two distinct loss functions. The likelihood function $L(\bm{\kappa} | \bm{S}, \sigma)$ is used to provide the gradient $\nabla_{\bm{\kappa}}$ to the network and is evaluated in the data input domain (i.e. weighted images). In contrast, Eq.\ref{eq6} is used to update the network parameters $\gamma$, and is evaluated in the parametric map domain (e.g. $T_1$ or $T_2$ relaxation maps). \par

A relevant feature of this framework is that the architecture of the RNNCell, more specifically, the number of input features in the first convolutional layer, only depends on $Q$, and not on $N$. This means the RIM can process series of weighted images $[\bm{S}_n]$ for $\forall N>0$.

\section{Methods}
\subsection{Sequences and parametric models}
The choice of parameters $\bm{\kappa}$ and the form of the parametric model $f_n$ depend on the pulse sequence used for acquisition. \par

For the $T_1$ mapping task in this work, we used the CINE sequence \citep{atkinson_edelman_1991}, based on a (popular) fast $T_1$ quantification method \citep{Look_1970}. It uses a non-selective adiabatic inversion pulse, applied after the cardiac trigger, with zero delay, and simulated at a constant rate of 100 beats per minute using a pulse generator developed in-house. For this sequence, a common parametric model is given by $f_n(\bm{\kappa}(\bm{x}_m))= \left | A \left(1-B  e^{-\frac{\tau_n}{T_1}}\right) \right |$, where $\tau_n$ is the $n^{th}$ inversion time and $\bm{\kappa}(\bm{x}_m)=(A, B, T_1)^T$ is the tissue parameter vector at position $\bm{x}_m$, in which $A$ is a quantity proportional to the proton density and receiver gain, $B$ is linked to the efficiency of the inversion pulse and $T_1$ is the longitudinal relaxation time. The operator $\mid \cdot \mid$ represents the element-wise modulus. \par
For $T_2$ experiments and quantification, we used the 3D CUBE Fast Spin-Echo sequence \citep{mugler_2014} with model given by $f_n(\bm{\kappa}(\bm{x}_m))= \left| A e^{-\frac{\tau_{n}}{T_2}} \right|$, where $\tau_n$ is the $n^{th}$ echo time and $\bm{\kappa}(\bm{x}_m)=(A, T_2)^T$, with $A$ proportional to the proton density and receiver gain and $T_2$ the transverse relaxation time.

\subsection{Generation of simulated data for training}
\label{section:training_data}

In this work, we opted to generate training data via model-based simulation pipeline. Training samples are composed of ground truth tissue parameters $\bm{\kappa}$ and their corresponding set of simulated weighted images $\bm{S}$. To generate training samples with a spatial distribution that resembles the human brain, ten 3D virtual brain models from the BrainWeb project \citep{Cocosco97brainweb:online} were selected. We randomly extract 2D patches from the brain models during training, with patch centers drawn uniformly from the model's brain mask. To introduce the notion of uniform tissue properties within subjects but distinct between subjects, for each patch and tissue separately, the parameters in $\bm{\kappa}$ were drawn from a normal distribution with values given in Table \ref{tab:tissue_param}. To enable recovery of intra-tissue variation, voxel-wise Gaussian noise was added to each parameter in $\bm{\kappa}$, except for $B$. Because the $B$ value is related to the efficiency of the inversion pulse in IR sequences, it is not tissue-specific, and as such, cannot be modeled as above. Its value was simulated as $2-\Gamma$, where $\Gamma$ is independently sampled, per patch, from the half-normal distribution \citep{leone_nelson_1961} with standard deviation $\sigma^{\Gamma}=0.2$. \par
Using $\bm{\kappa}$, $\bm{S}$ was simulated via Eq. (\ref{eq1}), with $\epsilon(\bm{x})$ an independent zero mean Gaussian noise where, for each patch, standard deviation $\sigma^{\text{acquisition}}$ was drawn from a log-uniform distribution with values in the range $[0.0065, $ $0.255]$, corresponding to SNR levels in the range of 100 to 3, respectively. \par

\begin{table}[hbt]
  \caption{Distribution of parameters per tissue and tissue property. $T_1$ and $T_2$ values in milliseconds. Values for $A$ are chosen as a fraction of the concentration of protons in the CSF.}
  \scriptsize
  \centering
  \begin{tabular}{m{1.25cm} m{0.75cm} m{0.75cm} m{0.75cm} m{0.75cm} m{0.75cm} m{0.75cm}}
   \toprule
    $\bm{\textbf{Tissue}}$ & $\mu_{\text{tissue}}^{T_1}$ & $\sigma_{\text{tissue}}^{T_1}$ & $\mu_{\text{tissue}}^{T_2}$ & $\sigma_{\text{tissue}}^{T_2}$ & $\mu_{\text{tissue}}^{A}$ & $\sigma_{\text{tissue}}^{A}$\\
   \midrule
    \textbf{CSF} 			& 3500 & 300 & 2000 & 300 & 1.0 & \multirow{18}{*} { \ \ 0.3} \\ \cmidrule{1-6}
    \textbf{GM} 			& 1400 & 300 & 110 & 30 & 0.85 &\\ \cmidrule{1-6}
    \textbf{WM} 			& 780 & 250 & 80 & 20 & 0.65 &\\ \cmidrule{1-6}
    \textbf{Fat} 			& 420 & 100 & 70 & 20 & 0.9 &\\ \cmidrule{1-6}
    \textbf{Muscle} 		& 1200 & 300 & 50 & 20 & 0.7 &\\ \cmidrule{1-6}
    \textbf{Muscle skin} 	& 1230 & 300 & 50 & 20 & 0.7 &\\ \cmidrule{1-6}
    \textbf{Skull} 			& 400 & 100 & 30 & 10 & 0.9 &\\ \cmidrule{1-6}
    \textbf{Vessels} 		& 1980 & 300 & 275 & 70 & 1.0 &\\ \cmidrule{1-6}
    \textbf{Connect.} 		& 900 & 250 & 80 & 20 & 0.7 &\\ \cmidrule{1-6}
    \textbf{Dura Mater} 	& 900 & 250 & 70 & 20 & 0.7 &\\ \cmidrule{1-6}
    \textbf{Marrow}		& 580 & 100 & 50 & 20 & 0.8 &\\ 
    \bottomrule
    \end{tabular}
\label{tab:tissue_param}
\end{table}

\begin{table*}[h!]
\setlength{\tabcolsep}{1.5pc}
\newlength{\digitwidth} \settowidth{\digitwidth}{\rm 0}
\catcode`?=\active \def?{\kern\digitwidth}
\newcolumntype{U}{>{\centering\arraybackslash}m{3.25cm}}
\newcolumntype{?}{!{\vrule width 0.15pt}}
\newcolumntype{@}{!{\vrule width 1.5pt}}
\Large
\centering
\aboverulesep=0ex
\belowrulesep=0ex
\renewcommand{\arraystretch}{1.25}
\caption{Acquisition settings for the evaluation datasets. $HP$ denotes the phantom scans while $IV$ are the in-vivo scans.}
\resizebox{\textwidth}{!}{\begin{tabular}{?p{5.5cm}@U?U?U?U?U?}
    \toprule
    Dataset & $\text{HP}_{T_1}$ & $\text{IV}_{T_1}$ & $\text{IV}^{\text{noisy}}_{T_1}$ & $\text{HP}_{T_2}$ & $\text{IV}_{T_2}$ \\ 
    \specialrule{.1em}{.1em}{.1em}
    FOV (pixel) & 210x210x15 & 210x210x10 & 210x210x1 & 210x210x15 & 210x210x10 \\ 
    \cmidrule{1-6}
    Slice thickness (mm) & 1.5 & 3.0 & 1.5 & 1.5 & 3.0 \\ 
    \cmidrule{1-6}
    Spacing (mm) & 1.5 & 1.5 & - & 1.5 & 1.5 \\ 
    \cmidrule{1-6}
    In-plane voxel size (mm) & \multicolumn{5}{c?}{0.82}  \\ 
    \cmidrule{1-6}
    Repetition Time (ms) & \multicolumn{3}{c}{8192} & \multicolumn{2}{?c?}{2010, 2020, 2040, 2080, 2160, 2320} \\ 
    \cmidrule{1-6}
    $\tau$ (Echo Times) (ms) & \multicolumn{3}{c}{4} & \multicolumn{2}{?c?}{10, 20, 40, 80, 160, 320} \\ 
    \cmidrule{1-6}
    $\tau$ (Inversion Times) (ms) & $\textbf{23 TIs}$: 172, 204, 237, 270, 303, 335, 368, 401, 434, 467, 499, 532, 565, 598, 630, 663, 696, 729, 761, 794, 827, 860, 893 & $\textbf{31 TIs}$: 139, 166, 193, 219, 246, 272, 299, 325, 352, 379, 405, 432, 458, 485, 511, 538, 565, 591, 618, 644, 671, 697, 724, 751, 777, 804, 838, 857, 883, 915, 937 & $\textbf{25 TIs}$: 172, 204, 237, 270, 303, 335, 368, 401, 434, 467, 499, 532, 565, 598, 630, 663, 696, 729, 761, 794, 827, 860, 893, 925, 958 & \multicolumn{2}{?c?}{-}\\ 
    \cmidrule{1-6}
    Flip Angle (º) & \multicolumn{3}{c}{10} & \multicolumn{2}{?c?}{-} \\ 
    \cmidrule{1-6}
    Acceleration factor & \multicolumn{5}{c?}{2} \\ 
    \cmidrule{1-6}
    C (nr. of repeated scans) & 4 & 2 & 2 & 4 & 2 \\ 
    \cmidrule{1-6}
    Acq. time/scan (min) & 4.3 & 7.5 & 1.6 & 3.2 & 3.2 \\ 
    \cmidrule{1-6}
\end{tabular}}
\label{tab:acquisition_params}
\end{table*}

\subsection{Evaluation datasets}

We performed all scans on a 3T General Electric Discovery MR750 clinical scanner (General Electric Medical Systems, Waukesha, Wisconsin) with a 32-channel head coil. 

\subsubsection{Hardware phantom} 
Phantom scans were carried out using the NIST/ISMRM system phantom \citep{keenan_2017} with parameters for the acquisition of $T_1$ weighted ($T_1w$) and $T_2$ weighted ($T_2w$) images presented in Table \ref{tab:acquisition_params} (datasets $\text{HP}_{T_1}$ and $\text{HP}_{T_2}$, respectively). The FOV contained the phantom’s $T_1$ array for $T_1w$ scans and the $T_2$ array for $T_2w$ scans. To evaluate the repeatability of each mapping method, $C=4$ consecutive acquisitions were performed without moving the phantom and with minimal time interval between scans.\par

\subsubsection{In-vivo} 
Our Institutional Review Board approved the volunteer study and informed consent was obtained from 2 healthy adults. $C=2$ repeated scans per volunteer were acquired for both $T_1$ and $T_2$ experiments to evaluate repeatability with in-vivo data. The FOV used was similar for $T_1$ and $T_2$ experiments and was oriented in the axial direction, with the middle slice positioned at the level of the body of the corpus callosum. These datasets, acquired with a slice thickness of 3mm, are referred to as $IV_{T_1}$ and $IV_{T_2}$, respectively. Details on acquisition settings are given in Table \ref{tab:acquisition_params}. Finally, to evaluate the performance of the estimators under low SNR conditions, we repeated the $T_1w$ acquisition using a slice thickness of 1.5mm (dataset called $IV^{\text{noisy}}_{T_1}$), in which a single slice, positioned above the corpus callosum, was acquired. Again, $C=2$ repeated scans were acquired for each volunteer to assess each method's repeatability.

\subsection{Implementation details}

The codes for all methods, trained models and the data used in the experiments are available online \footnote{https://gitlab.com/e.ribeirosabidussi/qmri-t1-t2mapping-rim}.

\subsubsection{MLE}

In the experiments in this study,  $\Psi(\bm{\kappa})$ is set as the sum over voxels of the voxel-wise square of the (spatial) Laplacian of $B$. A weighting term $\lambda_{B}$ is introduced to control the strength of the regularization and was empirically set to 500 to reduce the variability of the $T_1$ estimates. The remaining maps in the $T_1$ and $T_2$ mapping tasks are not regularized. \par

To prevent the estimator from getting stuck in a local minimum far from the optimal target, we initialize $\bm{\kappa}$ via an iterative linear search within a pre-specified range of values per parameter. Following initialization, parameters are estimated with a non-linear trust region optimization method. The estimation pipeline was implemented in MATLAB with in-house custom routines \citep{Poot_2015}. \par

\subsubsection{Network training}

To train both neural networks, 7200 2D patches of size $40\times40$ per brain model were generated during training and arranged in mini-batches of 24 samples, for a total of 3000 training iterations. \par

We used the ADAM optimizer with an initial learning rate of 0.001 and set the initial network weights with the Kaiming initialization \citep{Kaiming_2015}. PyTorch 1.3.1 was used to implement and train the models. The networks were trained on a GPU Nvidia P100, and all experiments (including timing) were performed on an Intel Core i5 2.7 GHz CPU.

\subsubsection{ResNet architecture}

Our implementation of the ResNet is a modified version of \cite{he_2016}. Pooling layers were removed to ensure limited influence between distant regions of the brain, effectively enforcing the use of local spatial context during inference. Additionally, our ResNet does not contain fully connected layers to adapt the network for a voxel-wise regression problem. All convolutions are zero-padded to maintain the patch size. \par

The first convolutional layer has a $1\times1$ filter, and it is used to increase the number of features from $N$ (the number of weighted images) to 40. This layer is followed by a batch normalization (BatchNorm) layer and a ReLu activation function. The core component of the network, denoted as the residual block (RB), comprises two $3\times3$ convolutional layers, two BatchNorm layers, and two ReLu activations, arranged as depicted on the right of Fig. \ref{fig:resnet_architecture}. Within a given RB, the number of features in each convolutional layer is the same. The skip connection is characterized by the element-wise addition between the input and the output of the second BatchNorm layer. In total, $\mathcal{G}=12$ residual blocks are sequentially linked, with number of feature channels in each block empirically chosen as $[40, 40, 80, 80, 160, 320, 160, 80, 80, 40, 6]$. The network architecture is completed by one $1\times1$ convolutional filter, used to reduce the number of features to $Q$.  Details on the general architecture are presented on the left of Fig. \ref{fig:resnet_architecture}.\par

Note that, due to differences in the inversion times used for the acquisition of $T_1$ weighted datasets (Table \ref{tab:acquisition_params}), we trained three ResNet models for the $T_1$ mapping task: (1) Training dataset generated with $N=23$ inversion times (ResNet$_{\boldsymbol{T_1 : 23}}$), (2) with $N=25$ inversion times (ResNet$_{\boldsymbol{T_1 : 25}})$, and (3) with $N=31$ inversion times (ResNet$_{\boldsymbol{T_1 : 31}}$). Finally, a fourth model was trained on the $T_2$ mapping task, denoted as ResNet$_{\bm{T_2}}$, with $N=6$ echo times. \par

\subsubsection{RIM architecture}

In this work, the RNNCell (shown in detail on the right of Fig. \ref{fig:rim_architecture}) is composed of four convolutional layers and 2 GRUs. The first $3\times3$ convolutional layer is followed by a hyperbolic tangent ($tanh$) link function, and its output, with 36 feature channels, is passed to the first GRU, which produces 36 output channels. The output of this unit ($\bm{h}^1_{j+1}$), also used as the first memory state, goes through two $3\times3$ convolutional layers with 36 output features, each followed by a $tanh$ activation. The data then passes through a second GRU, which generates the second memory state $\bm{h}^2_{j+1}$. The last layer is a $1\times1$ convolutional layer used to reduce the dimensionality of the feature channels, and it outputs $Q$ features, corresponding to the number of tissue parameters in $\bm{\kappa}$. All convolutional layers are zero-padded to retain the original image size. \par 

The parameter vector $\hat{\bm{\kappa}}$ was initialized as $A = \text{MIP}(\bm{S})$, $ B = 2$, $T_1 = 1000 \ ms$ and $T_2 = 100 \ ms$, where MIP is the Maximum Intensity Projection per voxel over all weighted images in the set. We used $J=6$ optimization steps for all RIM models. \par

Similarly to the ResNet, we trained three RIM models on the $T_1$ mapping ($RIM_{\boldsymbol{T_1 : 23}}$, $RIM_{\boldsymbol{T_1 : 25}}$, and $RIM_{\boldsymbol{T_1 : 31}}$) and one model on the $T_2$ task ($RIM_{\boldsymbol{T_2}}$). Notice that, while all $T_1$ datasets could be processed by a single RIM model, as the number of input features in the first convolutional layer does not depend on $N$, slight variations in inversion times might affect estimation error. This aspect will be assessed in Section \ref{section:experiments}, as it supplies information on the RIM's generalizability. \par

\subsection{Quantitative evaluation}
The prediction accuracy was evaluated in terms of the Relative Bias between the reference parameter values $\bm{\kappa}$ and the estimated parameters $\bm{\hat{\kappa}}^c \in \{\bm{\hat{\kappa}}^1, ..., \bm{\hat{\kappa}}^C\}$ for each repeated experiment $c$, defined as 
\begin{align}
\text{Relative Bias} \ [\%] = \frac{1}{C} \sum_{c=1}^{C} \left[\left(\bm{\hat{\kappa}}^{c} - \bm{\kappa}\right) \oslash \bm{\kappa}\right] \times 100\%, \label{eq9}
\end{align}
where $C$ is the number of repeated experiments and $\oslash$ denotes the element-wise division. The Coefficient of Variation (CV) was used to measure the repeatability of the predictions, and it is given by
\begin{align}
\text{CV} \ [\%]=   \left(\text{SD}^{c}\left(\bm{\hat{\kappa}}^{c}\right) \oslash \frac{1}{C}\sum_{c=1}^{C}\bm{\hat{\kappa}}^{c}\right) \times 100\%, \label{eq10}
\end{align}
where $\text{SD}^c$ denotes the standard deviation over $C$ estimates $\bm{\hat{\kappa}}$.

\section{Experiments}
\label{section:experiments}
\subsection{Simulated dataset}
\subsubsection{Noise robustness}
\label{noiseRobust}
To assess each method's robustness to noise and mapping quality, we generated the simulated $T_1w$ data with the process described in Section \ref{section:training_data} using a 2D slice of a virtual brain model not included in the training, matrix size $256\times256$ and inversion times of dataset $IV_{T_1}$. \par

For the same ground-truth $T_1, A \ \text{and} \ B$ maps, $C=100$ realisations of acquisition noise were simulated per $\text{SNR} \in [3,5,10,30,60,100]$. The Relative Bias and CV were computed per pixel and their distribution over all pixels within a brain mask is shown. The models RIM$_{\boldsymbol{T_1 : 31}}$ and ResNet$_{\boldsymbol{T_1 : 31}}$ were used in this experiment.

\subsubsection{Blurriness analysis}
We assessed the quality of the quantitative maps in terms of blurriness. Here, we defined blurriness as the amount of error introduced to a pixel, in terms of Relative Bias and CV, due to the influence of its neighbors and vice-versa. In this experiment, our interest lies on how well each mapping method can preserve the true $T_1$ value in small structures (e.g., one pixel), specifically hypo and hyper-intense regions that are at risk of being blurred away by the neural networks. \par
	To simulate the presence of these small anatomical structures, we changed the $T_1$ value of selected pixels in a ground-truth $T_1$ map (Fig. \ref{fig:result_blur}a), described as follows: $\Omega_{\text{point}}^{\text{hypo}}$ is a hypo-intense pixel ($T_1= 400ms$) within the gray mater of this map (shown in detail in Fig. \ref{fig:result_blur}b); $\Omega_{\text{point}}^{\text{hyper}}$ is a hyper-intense pixel ($T_1= 1200ms$) within the white mater (WM); $\Omega_{\text{line}}^{\text{vert}}$ is a hyper-intense vertical line ($T_1= 1200ms$) in the WM; and $\Omega_{\text{line}}^{\text{horz}}$ is a hyper-intense horizontal line ($T_1= 1200ms$) also in the WM. \par
	We measured the Relative Bias and CV per pixel in a Monte Carlo experiment with $C=100$ noise realizations (SNR=10). Each metric's median and standard deviation are reported for two disjoint regions in the estimated $T_1$ map, referred to as Structure and Neighborhood (Fig. \ref{fig:result_blur}c). This scenario, containing simulated structures is called $E2$, and was compared to the baseline error in the same regions in the original $T_1$ map (scenario $E1$). An independent t-test was applied to identify significant differences between $E1$ and $E2$. The models RIM$_{\boldsymbol{T_1 : 31}}$ and ResNet$_{\boldsymbol{T_1 : 31}}$ were used in this experiment.\par

\subsection{Evaluation with hardware phantom}
We manually drew ROIs within every sphere in the phantom and calculated the Relative Bias and CV per pixel within each ROI for $T_1$ and $T_2$ tasks. \par

Since nominal parameter values within the spheres, as reported in \cite{keenan_2017} and used as the reference $\bm{\kappa}$, include relaxation times shorter and longer than the $\tau$ used for training (Table  \ref{tab:acquisition_params}), we calculated the overall accuracy and repeatability as the average Relative Bias and CV over all pixels in spheres with parameter value in between the lowest and highest $\tau$. Because this dataset was acquired with 23 inversion times, models RIM$_{\boldsymbol{T_1 : 23}}$ and ResNet$_{\boldsymbol{T_1 : 23}}$ were used. \par

\subsection{Evaluation with In-vivo scans}
To evaluate the precision of estimates from in-vivo data, we compared $T_1$ and $T_2$ maps from all methods in terms of pixel-wise CV for all in-vivo scans. We also performed a visual comparison of the maps. \par
We evaluated the mapping quality in in-vivo scans regarding the sharpness of the boundary between gray mater and white mater. Twenty lines perpendicular to the tissue interface (Fig.  \ref{fig:sigmoid_fit_tissue_interface}a) were manually drawn in the measured quantitative maps. For each line, linear interpolation was used to reconstruct the $T_1$ values along them and a sigmoid model, given by $y(x)=V/(1+e^{-\upsilon \left(x-x_{0}\right)}) + b$, was fit using the MSE as objective function. The parameter $\upsilon$ denotes the slope of the fitted sigmoid and was used as a measure of boundary sharpness. A paired t-test was performed to evaluate significant differences between mapping methods.
	
\subsection{Model generalizability}

In this experiment, we evaluated how well the RIM can generalize to datasets with different acquisition settings, specifically, the variation of the inversion times in the three $T_1w$ datasets. In contrast to the ResNet architecture, which depends on the number of weighted images in the series, the RIM can process inputs of any length. \par
We used the three RIM$_{T_1}$ models (RIM$_{\boldsymbol{T_1 : 23}}$, RIM$_{\boldsymbol{T_1 : 25}}$ and RIM$_{\boldsymbol{T_1 : 31}}$) to infer $T_1$ maps from each $T_1w$ dataset, and computed the CV for the repeated experiments in each. The results were compared to the MLE and dataset-specific ResNet models.

\begin{figure}[ht!]
\centering
\includegraphics[trim={0.4cm 0.0cm 0.4cm 0.6cm}, clip, width=0.4\textwidth]{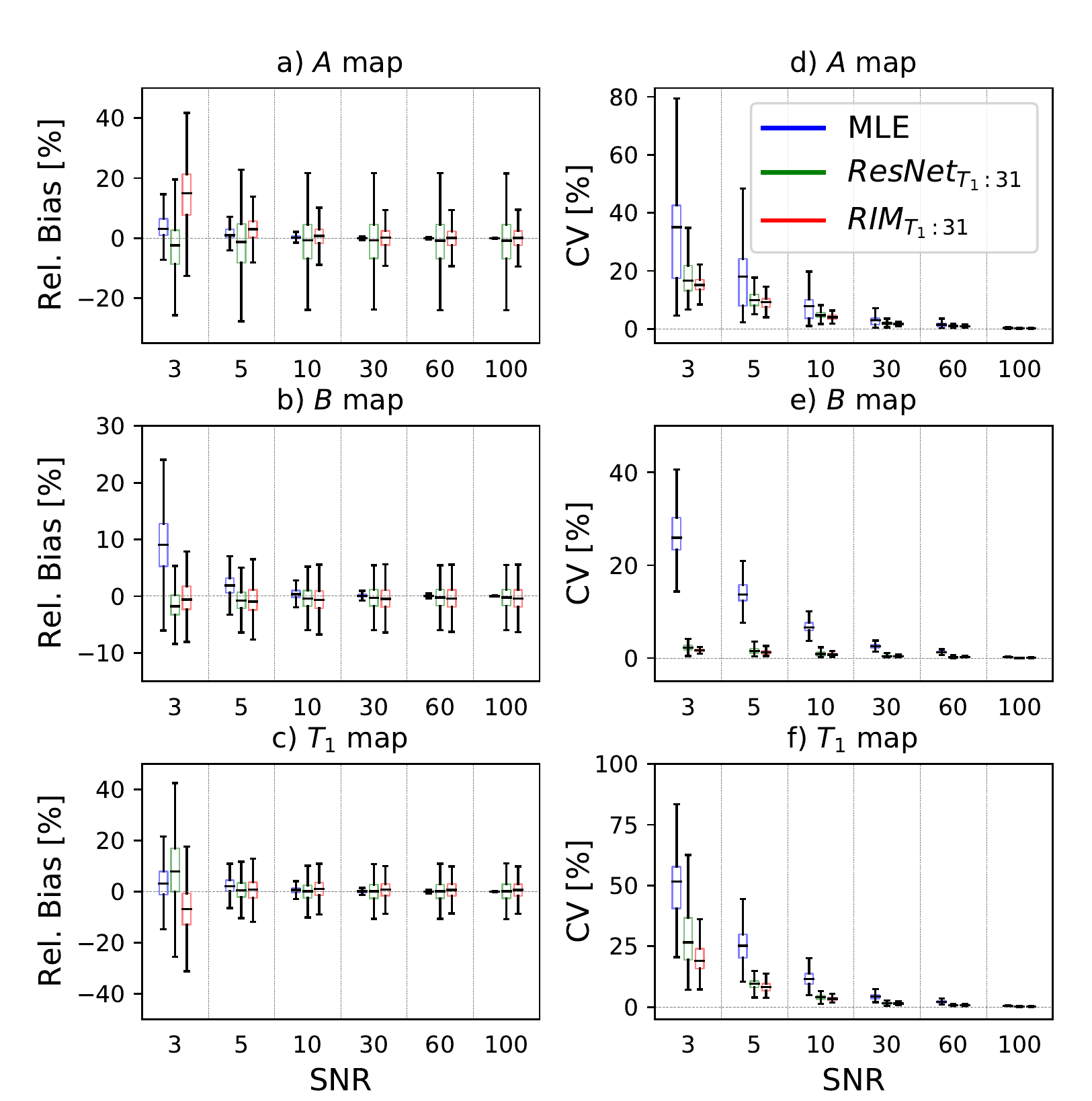} 
\caption{Results of the Monte Carlo experiment with a $T_1w$ simulated dataset for varying $SNR$ levels. a), b) and c) show the Relative Bias for the estimated $A$, $B$, and $T_1$ maps compared to simulated ground-truth. Figures d), e) and f) shows the Coefficient of Variation for the same maps. The boxplot represents the distribution of the metric over all pixels in the brain mask. The box extends from the lower to upper quartile values of this data, with a line at the median. The whiskers extend from the box to show the minimum and maximum values for each metric within the brain mask.}
\label{fig:noise_robustness_mc_simulation}
\end{figure}



\begin{figure*}[h!]
\centering
\includegraphics[trim={0cm 0.0cm 3cm 0cm}, clip, width=\textwidth]{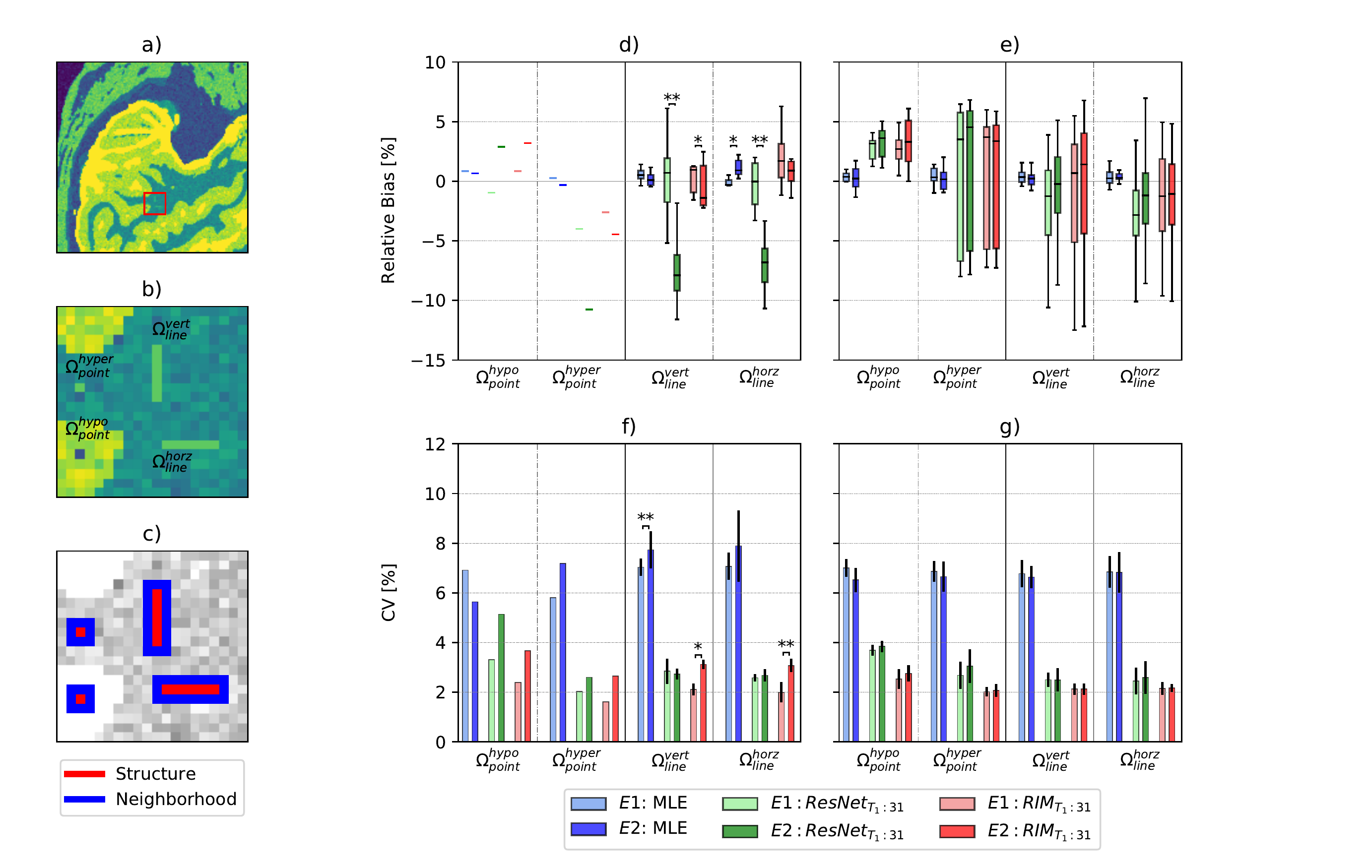} 
\caption{Evaluation of image blurriness in terms of Relative Bias and CV. a) Ground-truth $T_1$ map used to generate the weighted images $S$. The red box indicates the position of the simulated artefacts. b) The four simulated structures. c) Representation of the areas of interest. The blue areas are the structures, and red areas are their immediate neighborhood. d) Relative Bias over one hundred repetitions within the Structure region. e) Relative Bias over one hundred repetitions within the Neighborhood region. f) CV over one hundred repetitions within the Structure region. g) CV over one hundred repetitions within the Neighborhood region. In all plots, the box extends from the lower to upper quartile values of the data, with a line at the median. The whiskers extend from the box to show the range of the data. The vertical black lines at the top of the bars (plots f) and g)) show the standard deviation over the data. Significant differences between scenarios $E1$ and $E2$ are indicated by $^{*}$ and $^{**}$, representing $p<0.05$  and $p<0.01$, respectively.}
\label{fig:result_blur}
\end{figure*}


\section{Results}
\subsection{Simulated dataset}
Figures \ref{fig:noise_robustness_mc_simulation}(a)-(c) show the Relative Bias measured for $A$, $B$ and $T_1$ maps in the experiment with simulated $T_1w$ data. For most cases where SNR $>3$, all methods produced quantitative maps with comparable median Relative Bias, but both neural networks displayed a larger range of values than the MLE. The CV for all SNR levels is shown in Figs. \ref{fig:noise_robustness_mc_simulation}(d)-(f) for the same data. The RIM presented lower CV than the other methods for all SNRs. In comparison, the MLE displayed significantly higher CV compared to RIM and ResNet, accentuated in low SNR. The results of the experiments with simulated $T_2w$ data were similar and are shown in Fig. A1 of the Supplementary Results. \par

Figures \ref{fig:result_blur}(d)-(g) show the results of the blurriness analysis. Specifically, Figs. \ref{fig:result_blur}(d) and \ref{fig:result_blur}(f)  depict the Relative Bias and CV measured per pixel within the Structure area. We observe that both neural networks presented increased Relative Bias compared to scenario $E1$. For the RIM, the highest increase occurred for $\Omega_{\text{point}}^{\text{hypo}}$, with Relative Bias going from $0.68\%$ to $3.43\%$. This difference represents an average error of 11ms over the ground-truth $T_1$ value of $400ms$, or a loss of $0.81\%$ in $T_1$ contrast between the pixel and its neighbors, with average $T_1$ of $1350ms$. The ResNet showed considerably higher bias than RIM when small structures were added, while for the MLE, the difference between scenarios $E1$ and $E2$ is not significant (with exception for $\Omega_{\text{line}}^{\text{horz}}$). The RIM showed increased CV for all structures compared to the baseline, but values were still lower than the MLE's and comparable to the ResNet's. Figures  \ref{fig:result_blur}(e) and \ref{fig:result_blur}(g) show the Relative Bias and CV for the Neighborhood region. We observe higher Relative Bias for RIM and ResNet than the MLE, with a wider range of values, but we found no significant differences between $E1$ and $E2$ for any of the cases. \par

The average computing time to produce $\bm{\hat{\kappa}}$ from $N=31$ weighted images (with size 256 $\times$ 256 pixels) was measured as 3.8$s$ for the $\text{RIM}_{T_1:31}$, 27$s$ for $\text{ResNet}_{T_1:31}$ and 575$s$ for the MLE.

\begin{figure}[htb]
\centering
\includegraphics[trim={0cm 0cm 0cm 0cm}, clip, width=0.7\linewidth]{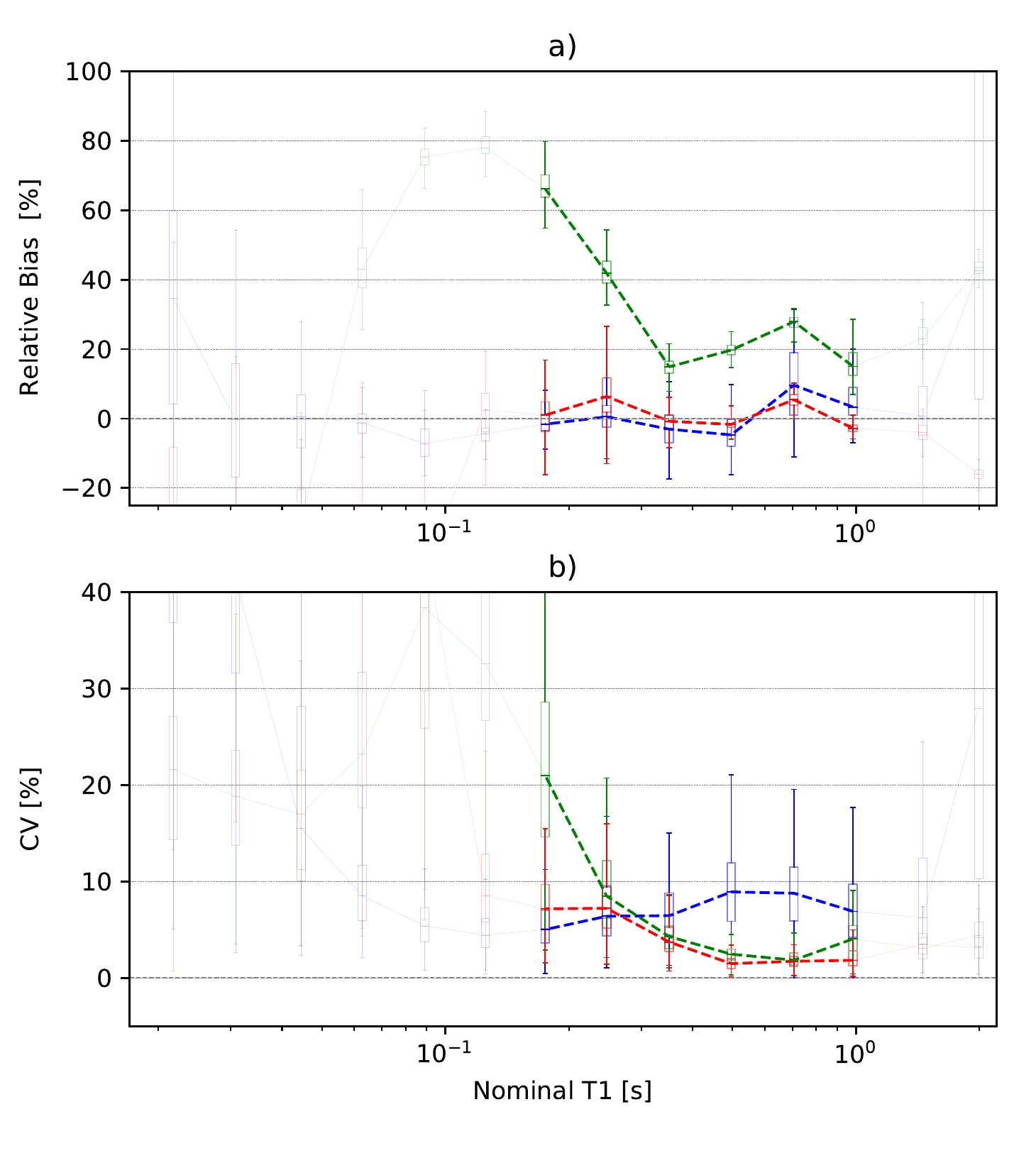}
\caption{Estimation of $T_1$ values in the ISMRM/NIST phantom. a) Distribution of Relative Bias over all pixels within a ROI versus nominal $T_1$ values in the phantom. b) Box plot of the CV in the different spheres/ROIs of the phantom, plotted as a function of their nominal $T_1$ value. In both figures, the fully-coloured strokes indicate the spheres with $T_1$ values within the range of inversion times.}
\label{fig:phantom_analysis}

\end{figure}


\subsection{Evaluation with hardware phantom}
The $T_1$ quantification results are shown in Fig. \ref{fig:phantom_analysis}. In Fig. \ref{fig:phantom_analysis}(a) we present the Relative Bias for the different spheres in the phantom. The average Relative Bias was computed over the spheres in the restricted $\tau$ domain (full-color lines), in which the RIM$_{\boldsymbol{T_1 : 23}}$ model shows lower error (1.34$\%$) compared to the MLE (1.71$\%$) and ResNet$_{\boldsymbol{T_1 : 23}}$ (31.06$\%$). The CV as a function of $T_1$ values is shown in Fig. \ref{fig:phantom_analysis}(b). The average CV over the restricted $\tau$ domain was measured as 3.21$\%$ for RIM$_{\boldsymbol{T_1 : 23}}$,  7.56$\%$ for MLE and 7.5$\%$ for ResNet$_{\boldsymbol{T_1 : 23}}$. \par
	The results for the $T_2$ mapping task with the hardware phantom are shown in Fig. A2 of the Supplementary Results, where we observed larger Relative Bias for all methods. \par
	
\begin{figure}[!t]
\centering
\includegraphics[width=0.9\linewidth]{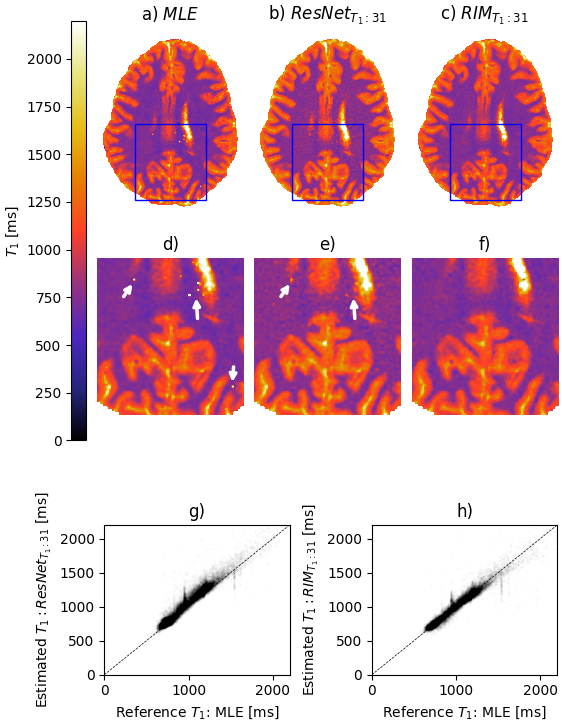}
\label{fig:In_vivo_3mm_maps}\caption{$T_1$ maps estimated from the $\bm{IV_{T_1}}$ dataset. Scan 1 of volunteer 1 is shown. a-c) $T_1$ maps generated by each mapping method and the detail (blue box) shown in figures d-e). The white arrows indicate estimation outliers. g) Agreement between the ResNet and MLE and h) RIM and MLE.}
\label{fig:t1_mapping_3mm}
\end{figure}

\begin{figure}[!t]
\centering
\includegraphics[width=0.99\linewidth]{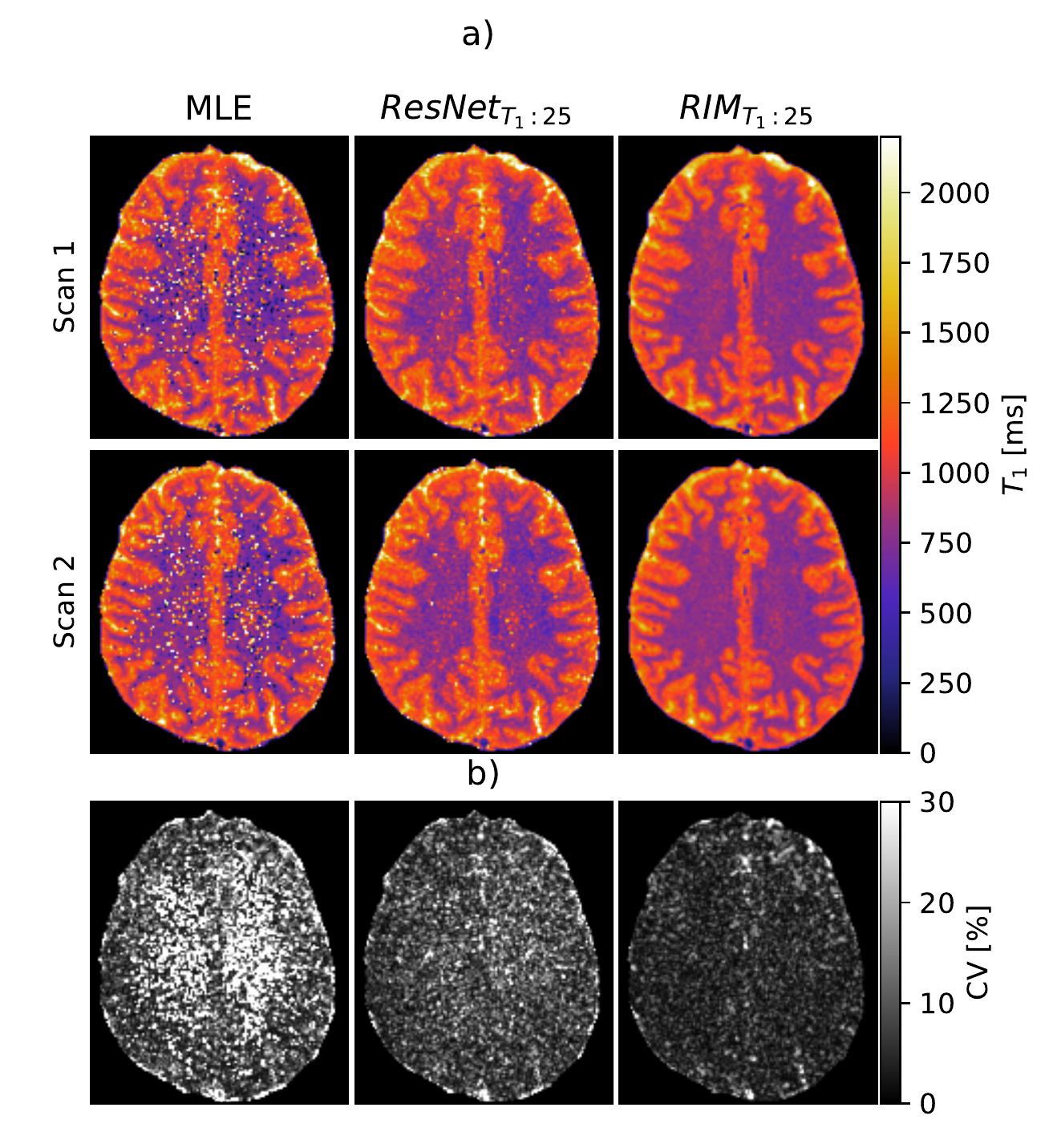}
\caption{$T_1$ maps estimated from the $IV_{T_1}^{noisy}$  dataset a) $T_1$ maps estimated from volunteer 1 for repeated scans 1 and 2. b) Their respective pixel-wise CV map.}
\label{fig:t1_mapping_1mm}
\end{figure}

\subsection{Evaluation with In-vivo scans}
The $T_1$ maps generated by each method for volunteer 1 in the low noise dataset $IV_{T_1}$ are shown in Figs. \ref{fig:t1_mapping_3mm}(a)-(c). We observe the presence of outliers in the MLE and ResNet$_{\boldsymbol{T_1 : 31}}$ (white arrows in Figs. \ref{fig:t1_mapping_3mm}(d)-(f)), while the RIM$_{\boldsymbol{T_1 : 31}}$ produced a clean $T_1$ map. The scatter plot in Fig. \ref{fig:t1_mapping_3mm}(h) shows that the RIM estimate is nearly unbiased when compared to the MLE's, while the ResNet presented overestimated $T_1$ values (Fig. \ref{fig:t1_mapping_3mm}(g)). \par

$T_1$ maps inferred from the noisier dataset $IV^{\text{noisy}}_{T_1}$ are shown in Fig. \ref{fig:t1_mapping_1mm}(a). The RIM$_{\boldsymbol{T_1 : 25}}$ showed increased noise robustness compared to the MLE and ResNet$_{\boldsymbol{T_1 : 25}}$, clearly outperforming these methods in terms of outliers. The CV maps, computed per pixel, are presented in Fig. \ref{fig:t1_mapping_1mm}(b) and shows that the RIM$_{\boldsymbol{T_1 : 25}}$ model produces low-variance quantitative maps, with average CV over all pixels equal to $6.4\%$, compared to $17.1 \%$ from the MLE and $11.06 \%$ from the ResNet$_{\boldsymbol{T_1 : 25}}$. \par
\begin{figure}[h!]
\centering
\includegraphics[trim={0cm 0.05cm 1.5cm 0cm}, clip, width=0.8\linewidth]{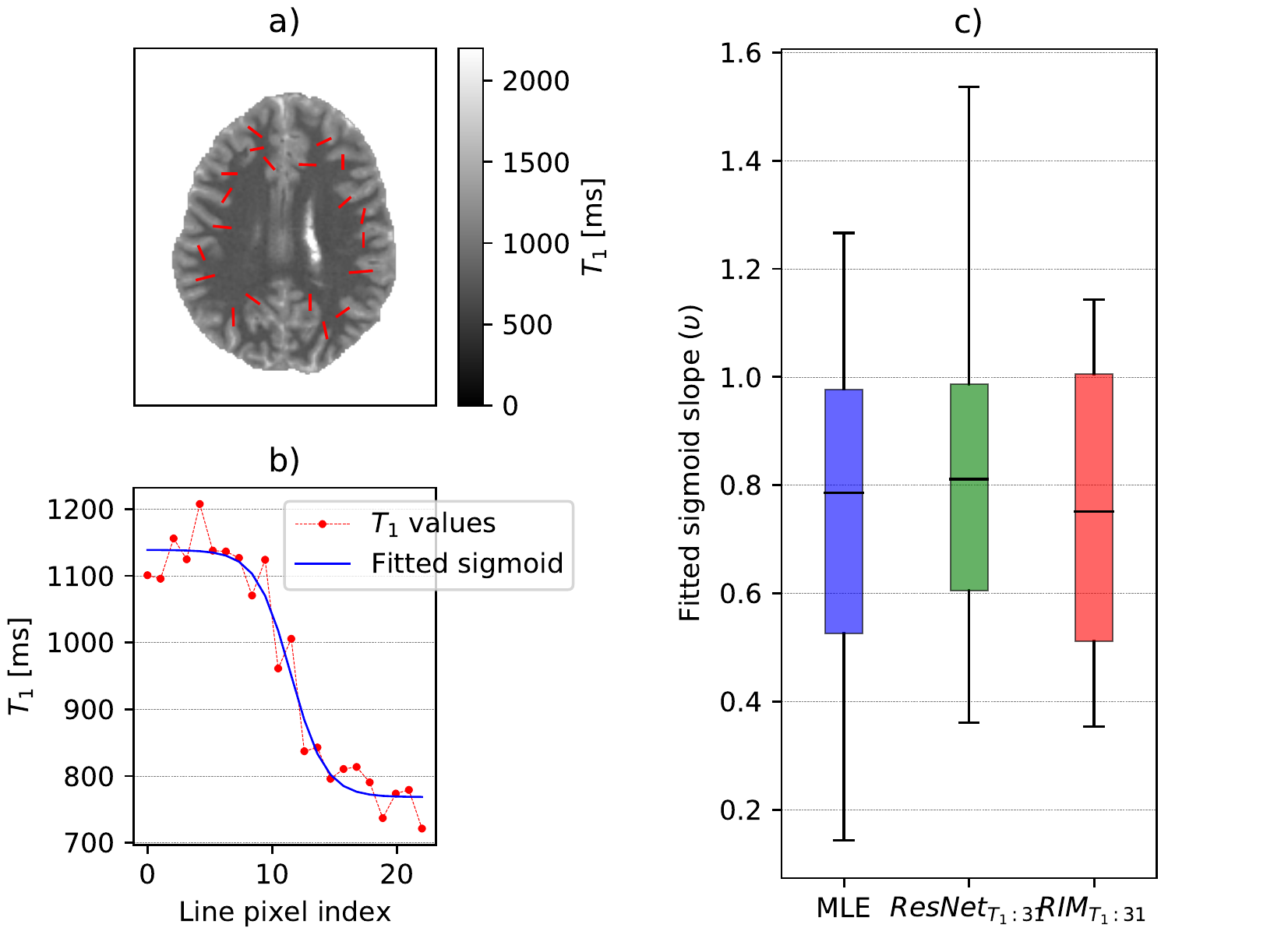}\label{fig:in_vivo_sigmoid_detail}
\caption{Evaluation of the integrity of the GM/WM boundaries. a) Detail on the twenty lines were manually drawn perpendicular to the GM/WM interface indicated by the red lines. b) An example of the sigmoid fitting for one of the lines. c) The box plot depicts distribution of the absolute sigmoid slope ($\upsilon$) for all 20 lines for each mapping method. We found no significant differences between the methods.}
\label{fig:sigmoid_fit_tissue_interface}
\end{figure}

Figure \ref{fig:sigmoid_fit_tissue_interface}(c) shows the result of the image quality analysis for in-vivo scans. The figure depicts the distribution of the sigmoid slope $k$ for each method across all 20 lines. The whiskers indicate the minimum and maximum $k$ values, the boxes show the lower and upper quartiles and the solid horizontal line their median. The paired t-test shows no significant differences between methods. \par
\begin{figure}[!t]
\centering
\includegraphics[trim={7cm 1.3cm 8.7cm 0.5cm}, clip, width=0.75\linewidth]{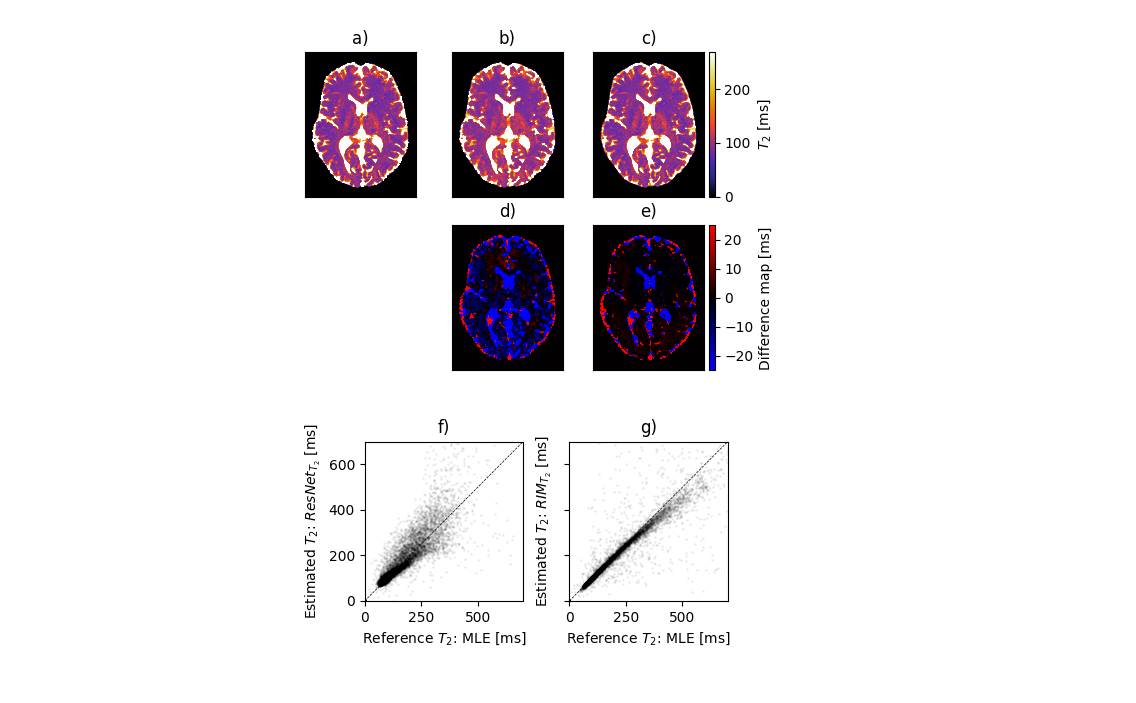}
\label{fig:t2_maps}
\caption{$T_2$ estimation. a-c) (left to right) $T_2$ maps from dataset $IV_{T_2}$ for the MLE, ResNet and RIM. d) Difference map between the ResNet and MLE and e) RIM and MLE. f-g) Scatter plots showing the agreement between each neural network and the reference MLE.}
\label{fig:t2_mapping}
\end{figure}

Figures \ref{fig:t2_mapping}(a)-(c) show the $T_2$ maps generated by each mapping method. The RIM$_{\boldsymbol{T_2}}$ predicted $T_2$ values that are similar to the reference MLE, with average difference in $T_2$ of $-1.13ms$ across all pixels in the brain, while the ResNet$_{\boldsymbol{T_2}}$ again showed overestimated relaxation times compared to the MLE, with an average difference of $26.2ms$. Difference maps between the MLE and both neural networks are shown in Figs. \ref{fig:t2_mapping}(d) and \ref{fig:t2_mapping}(e). The scatter plots in Figs. \ref{fig:t2_mapping}(f) and \ref{fig:t2_mapping}(g) depict the agreement between the neural network estimates and the reference MLE.

\subsection{Model generalizability}
\begin{figure}[!t]
\centering
\includegraphics[trim={0.4cm 0cm 0.7cm 0.5cm}, clip, width=0.45\textwidth]{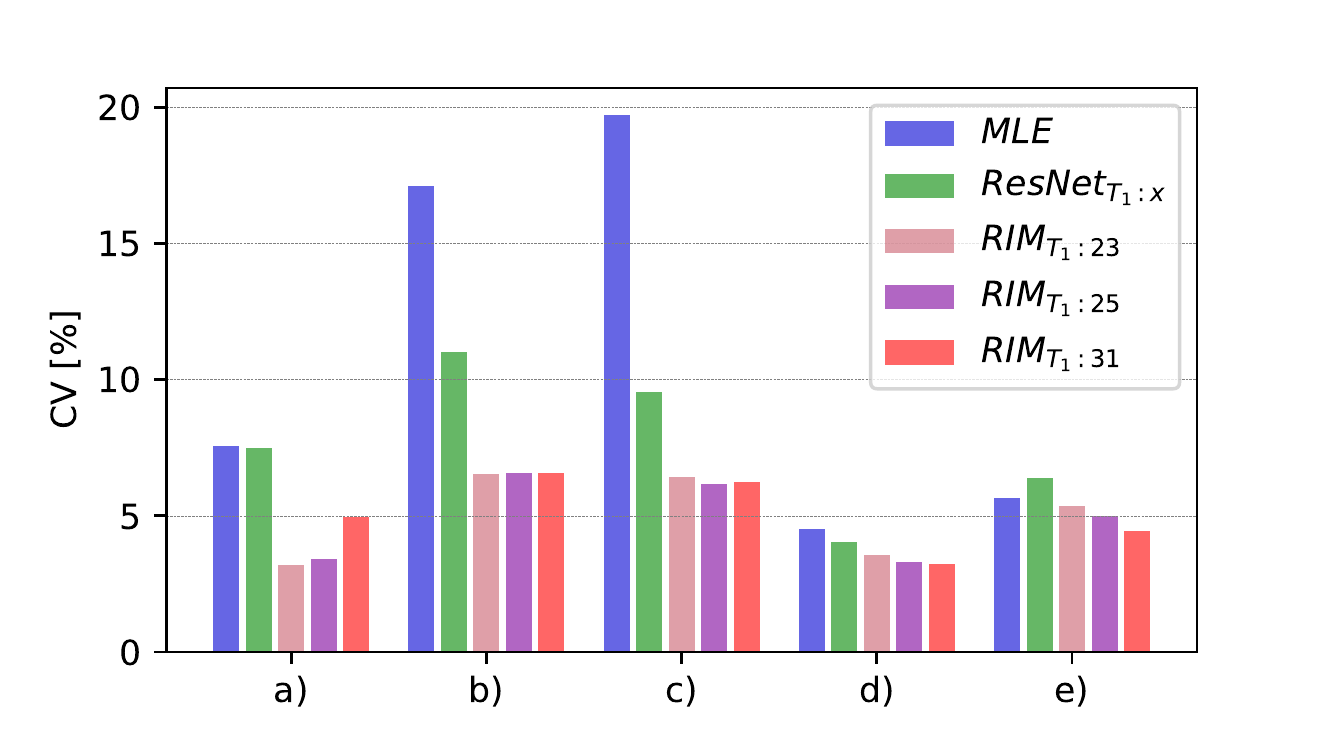} 
\caption{Results of the model generalisability experiment. The 3 RIM models ($\text{RIM}_{T_1:23}$, $\text{RIM}_{T_1:25}$ and $\text{RIM}_{T_1:31}$) for $T_1$ mapping were used to estimate data from all datasets and compared to the results from MLE and ResNet. a) Dataset $\bm{HP_{T_1}}$ (23 TIs), b) dataset $\bm{IV^{noisy}_{T_1}: Volunt.1}$ (25 TIs) c) dataset $\bm{IV^{noisy}_{T_1}: Volunt.2}$ (25 TIs) d) dataset $\bm{IV_{T_1}: Volunt.1}$ (31 TIs) e) $\bm{IV_{T_1}: Volunt.2}$ (31 TIs). The median CV over all pixels containing tissues of interest (phantom spheres or brain tissue) is shown.}
\label{fig:T1_generalization}
\end{figure}

Fig. \ref{fig:T1_generalization} illustrates the CV of the different models evaluated on all $T_1w$ datasets. The graph shows that the RIM produces estimates with lower variance than the MLE and ResNet, regardless of the number of inversion times used to create the training set. Note that, in every case, the RIM trained for the specific data performs slightly better than the other RIM models. However, we found no significant differences in repeatability between these models.

\section{Discussion}

This work presented a novel approach for MR relaxometry using Recurrent Inference Machines. Previous works showed that RIMs produce state-of-the-art predictions solving linear reconstruction problems. Here, we expanded the framework and demonstrated that it could be successfully applied to non-linear inference problems, outperforming a state-of-the-art Maximum Likelihood Estimator and a ResNet model in $T_1$ and $T_2$ mapping tasks. \par
	In simulated experiments, we observed that the RIM reduces the variance of estimates without compromising accuracy, suggesting higher robustness to acquisition noise than the MLE, and attesting to the advantages of using the neighborhood context in the inference process. In addition, for low SNR, the RIM had lower variance than the ResNet, suggesting that the neighborhood context alone is not the sole responsible for the increased quality, and that the data consistency term (likelihood function) in the RIM framework helps to produce more reliable estimates. This showcases a major advantage of the RIM framework over current conventional and deep learning methods for QMRI. \par	
	The phantom experiments performed to assess the Relative Bias and CV in real, controlled scans showed that the RIM has the lowest Relative Bias among the evaluated methods. The ResNet presented significantly higher error, which indicates that the ResNet does not generalize well to unseen structures, and the use of simulated training data with this model should be carefully considered. Because the RIM can generalize well, using simulated data for training represents a significant advantage over models trained with real-data when considering dataset flexibility, since any combination of parameter values can be simulated and the training dataset can be arbitrarily large.\par
	In all in-vivo scans, the RIM produces quantitative maps similar to those from the MLE, with higher robustness to noise. Although the ResNet estimates parametric maps consistent with reported $T_1$ and $T_2$ relaxation times of brain tissues, they are often overestimated compared to the MLE. In terms of coefficient of variation, the RIM results are superior compared to the other methods, independently of the dataset.  \par
	The anatomical integrity of quantitative maps is an essential factor when evaluating the quality of a mapping method. The RIM and the ResNet use the pixel neighborhood's information to infer the parameter value at that pixel, which creates valid concern regarding the amount of blur introduced by the convolutional kernels. We demonstrated in simulation experiments that, although the RIM does introduce a limited amount of blur to the quantitative maps, small structures are still confidently retained, and the error introduced by the pixel neighborhood does not represent a significant change in the relaxation time of those structures. Additionally, in in-vivo experiments, both deep learning methods produce relaxation maps with similar structural characteristics to the maps inferred by the MLE. More concretely, the $T_1$ relaxation times in the interface between gray and white mater follow a similar transition pattern to the MLE, further suggesting that the RIM does not introduce sufficient blur to alter brain structures, even in in-vivo scans. \par

\section{Conclusion}
We proposed a new method for $T_1$ and $T_2$ mapping based on the Recurrent Inference Machines framework. We demonstrated that our method has higher precision than, and similar accuracy levels as an Maximum Likelihood Estimator and higher precision and higher accuracy than an implementation of the ResNet. The experimental results show that the proposed RIM can generalize well to unseen data, even when acquisition settings vary slightly. This allows the use of simulated data for training, representing a substantial improvement over previously proposed QMRI methods that depend on alternative mapping methods to generate ground-truth labels. Lastly, the RIM dramatically reduces the time required to infer quantitative maps by 150-fold compared to our implementation of the MLE, showing that our proposed method can be used in large studies with modest computing costs.\par

\section*{Acknowledgements}
This work is part of the project B-QMINDED which has received funding from the European Union's Horizon 2020 research and innovation programme under the Marie Sklodowska-Curie grant agreement No 764513.

\biboptions{authoryear}
\bibliographystyle{elsarticle-harv}
\bibliography{mapping_rim}

\end{document}